\documentstyle[prb,aps,psfig]{revtex}
\newcommand \be{\begin{eqnarray}}
\newcommand \ee{\end{eqnarray}}
\begin{document}
\twocolumn[\hsize\textwidth\columnwidth\hsize
           \csname @twocolumnfalse\endcsname
\title{Bernoulli potential in type-I and weak type-II 
superconductors: Surface charge}
\author{P. Lipavsk\'y$^1$, K. Morawetz$^{2,3}$, 
J. Kol\'a\v cek$^1$, 
J. J. Mare{\v s}$^1$, E. H. Brandt$^4$, M. Schreiber$^2$}
\address{$^1$Institute of Physics, Academy of Sciences, 
Cukrovarnick\'a 10, 16253 Prague 6, Czech Republic\\
$^2$Institute of Physics, Chemnitz University of Technology, 
09107 Chemnitz, Germany\\
$^3$Max-Planck-Institute for the Physics of Complex
Systems, Noethnitzer Str. 38, 01187 Dresden, Germany\\
$^4$Max-Planck-Institute for Metal Research,
         D-70506 Stuttgart, Germany}
\maketitle
\begin{abstract}
The electrostatic potential close to the surface of 
superconductors in the Meissner state is discussed. We show
that beside the Bernoulli potential, the quasiparticle 
screening, and the thermodynamic contribution due to 
Rickayzen, there is a non-local contribution which is
large for both type-I and weak type-II superconductors. 
\end{abstract}
    \vskip2pc]

\section{Historical review}
The electrostatic potential in superconductors is known already
for seven decades. Within the pre-London approaches based on the 
ideal charged liquid, the electrostatic potential was assumed to 
balance the Lorentz and the inertial forces acting on diamagnetic 
currents.\cite{B37} The current ${\bf j}=en{\bf v}$ flows along 
the surface and its amplitude falls off exponentially from the 
surface into the bulk. Similarly the screened magnetic 
field decays. The balancing electrostatic potential was found to be of 
the Bernoulli type, $e\varphi=-{1\over 2}mv^2$, therefore it 
falls off on the scale of half the London penetration depth. 
From the Poisson equation, 
\begin{equation}
-\epsilon_0\nabla^2\varphi=\rho,
\label{a4}
\end{equation}
one finds that this corresponds to a charge accumulated in the 
layer penetrated by the magnetic field. To maintain the charge 
neutrality, the accumulated bulk charge has to be accompanied
by the opposite surface charge. This picture has been 
confirmed by the London theory.\cite{L50}

\subsection{Advanced London-type approaches}
The electrostatic potential equals the Bernoulli potential 
only at zero temperature when all electrons are in the 
superconducting condensate. At finite temperatures, a part 
$n_n$ of electrons remains in the normal state, while the rest 
$n_s=n-n_n$ contributes to the supercurrent, ${\bf j}=en_s
{\bf v}$. The electric force acting on normal electrons is
transferred to the condensate via a mechanism which reminds one of
the fountain effect in superfluid helium. As a consequence, 
the electrostatic potential reduces to $e\varphi=-{n_s\over n}
{1\over 2}mv^2$.  This reduction of the potential derived by 
van~Vijfeijken and Staas\cite{VS64} has a somehow confusing 
name: the quasiparticle screening. It should be noted 
that the electrostatic potential caused by currents in 
superconductors is traditionally called the Bernoulli 
potential in spite of the quasiparticle screening and other 
non-Bernoulli contributions found later. 

Jakeman and Pike recovered the result of van~Vijfeijken
and Staas from the static and the classical limit of the 
time-dependent Ginzburg-Landau (GL) theory.\cite{JP67} 
Unlike previous studies, their approach includes the real 
screening due to spatial distribution of the charge, 
therefore it describes the surface charge as a space charge 
localized on the scale of the Thomas-Fermi screening length. This picture 
of the surface charge was accepted for a long time as it looks 
natural and seems to be justified from the semi-microscopic 
theory. 

Later studies of the electrostatic potential did not re-address
the question of the surface charge but attempted to overcome
the hydrodynamic picture common to all the above mentioned 
studies. Naturally, they focused on a detailed description of the 
Bernoulli potential on the scale of the London penetration 
depth. 

Adkins and Waldram recovered the Bernoulli potential from the 
BCS theory for the system at zero temperature.\cite{AW68} 
They also indicated that at finite temperatures one should 
expect a contribution which depends on the band structure. 

The predicted contribution has been evaluated by 
Rickayzen.\cite{R69} He used the thermodynamic approach which
yields 
\be
e\varphi=-{\partial n_s\over\partial n}\, \frac 1 2 mv^2.
\ee 
The
density dependence of the condensate density can be evaluated 
either from the BCS theory leading to the result of Adkins and 
Waldram or from the two-fluid relation $n_s\approx n\left(1-
{T^4\over T_c^4}\right)$. The latter gives a simple 
formula 
\be
e\varphi=-{n_s\over n}{1\over 2}mv^2-4{n_n\over n}
{\partial\ln T_c \over\partial\ln n}{1\over 2}mv^2.
\label{waldram}
\ee 
The first term balances the Lorentz and inertial forces, while 
the second reflects the pairing. We will call terms 
proportional to ${\partial\ln T_c\over\partial\ln n}$ the 
thermodynamic correction.

In the next decade, the interest in the electrostatic potential 
in superconductors shifted towards non-equilibrium situations, 
see e.g. Ref.~\onlinecite{G81}. It should be noted that the 
Bernoulli potential introduced within the theory of 
non-equilibrium systems is not the same as the one we discuss.
Within the non-equilibrium theory the term Bernoulli potential 
is used for the velocity-dependent part of the energy of Cooper
pairs. Only in equilibrium, the velocity-dependent energy 
agrees with the electrostatic potential we discuss here.

\subsection{GL approaches -- local}
New studies of the Bernoulli potential in equilibrium have 
been stimulated by the question of the vortex charge. Khomskii 
{\em et al.}\,\cite{KK92,KF95} have used the idea of 
van~der~Marel\cite{M89} to derive a simple estimate of the
electrostatic potential in terms of the superconducting gap.
The obtained formula corresponds to the thermodynamic
correction of Rickayzen with the original balance term being omitted.
While the validity of the approach is restricted to temperatures
close to $T_c$, it has the advantage, however, that it can be 
applied within the GL theory. Blatter {\em et al.}\,\cite{B96} 
have employed this estimate to predict a possible electrostatic 
field above the surface of the superconductor with the vortex 
lattice perpendicular to it. 

An opposite limit represents the treatment of the vortex charge 
by LeBlanc.\cite{L97} He considers only the balance term 
omitting the thermodynamic correction completely. As one can 
see from Rickayzen's formula, this approximation is limited to 
the region of low temperatures.

\subsection{GL approaches -- non-local}
All the above mentioned theories have in common that the
electrostatic potential is a local function of the classical 
kinetic energy or of the gap. In other words, they neglect the
gradient of the condensate. The non-local corrections within
the GL theory have been proposed in Ref.~\onlinecite{KLB01}.
It turned out that the non-local corrections are taken into account 
if the classical kinetic energy in the balance equation is replaced 
by its quantum-mechanical counterpart. 

For our next discussion it is important that the non-local 
contributions can be rearranged into a local form.\cite{KLB01}
Indeed, within the GL theory the sum of the kinetic energy  and the
GL potential is zero which allows us to express the balance 
term via the local GL potential. Note that this rearrangement
is possible only within the non-local theory, because 
gradients of the condensate density $n_s=2|\psi|^2$, where $\psi$ is the GL wave function, provide
important contributions to the quantum kinetic energy, in particular
close to the surface.

In Ref.~\onlinecite{KLB01}, the quasiparticle screening and 
the thermodynamic correction have been neglected, the complete 
theory has been presented in Ref.~\onlinecite{KL01}. 
The detailed derivation together with estimates of material 
parameters and the numerical treatment of the Abrikosov vortex 
lattice in niobium at various temperatures can be found in 
Ref.~\onlinecite{LKMB02}. Again, all contributions including
the kinetic energy can be expressed in a local relation
$\varphi[|\psi|^2]$.

The non-local form of the Bernoulli potential requires to
reconsider the picture of surface charges. It is paradoxical 
that the possibility to rearrange the non-local corrections
into a local function plays the crucial role. As noticed 
already by Yampolskii {\em et al.},\cite{YBPK01} if the 
electrostatic potential is of the form $\varphi[|\psi|^2]$, the
GL boundary condition, $\nabla\psi=0$ in the direction normal
to the surface, implies that no surface charge is needed.
Indeed, at the surface $\nabla\varphi={\partial\varphi\over
\partial\psi}\nabla\psi=0$. The electric field thus vanishes
at the surface which is a sufficient condition for the charge
neutrality. 

\subsection{Plan of the paper}
Apparently, there is no surface charge on the scale of the
Thomas-Fermi screening length. The electric field due to the 
dipole between the surface charge and the bulk charge was 
supposed to balance the Lorentz force, therefore one has to 
ask the question how the balance of forces looks if the surface
charge is absent.

The charge profile in the vicinity of the surface has been 
numerically studied in Ref.~\onlinecite{KLB02}. This study
was limited to thin slabs and type-II superconductors, its
results indicate, however, that the charge profile at the surface
tends to organize into two layers - a narrow surface charge
and a thicker bulk charge. 

In this paper we discuss the charge and potential profile
at the surface of a semi-infinite superconductor in the 
Meissner state. In the next section we shortly describe the surface 
charge within the GL theory. In section III we discuss the limit of
a weak magnetic field, where all results can be derived
analytically. We will show that the Bernoulli potential at
the surface differs from the value predicted by Rickayzen. The
correction has the  form of a multiplicative factor that approaches 
unity for the extreme type-II superconductor. Numerical results on strong 
magnetic fields are presented in section IV revealing a nonlinear mechanism 
which, however, is found to be very small. The conclusions and outlook are 
presented in section V.

\section{Surface charge within the Ginzburg-Landau type 
theory}
\begin{figure}[h]  % F1
\psfig{figure=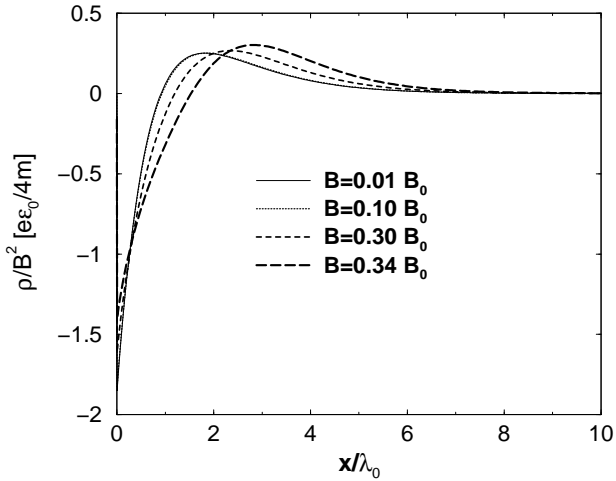,width=8cm,angle=-90}
\caption{The charge profile close to the surface for 4 different magnetic fields. 
The magnetic field is given in dimensionless units proportional to the magnetic flux quanta, $B_0= 
\Phi_0/(2\pi\lambda_0^2)=e\mu_0 n \hbar/2m$. The coordinate $x$ 
is measured in units of the London penetration depth $\lambda_0$ at 
zero temperature. The assumed temperature $t=T/T_c=0.9$ yields
the penetration depth $\lambda=1.7\lambda_0$. The 
charge is scaled with $B^2$ so that the two lowest magnetic 
fields $B=0.01 B_0$ and $B=0.1 B_0$ give nearly identical curves. 
The magnetic field $B=0.34 B_0$ is close to the critical value.
The GL parameter $\kappa_{\rm GL}
=1$ corresponds to weak type-II superconductors. 
}
\label{f1}
\end{figure}

A typical profile with the charge distribution close to the
planar surface and homogeneous external magnetic fields is 
shown in Fig.~\ref{f1}. The profile has been obtained by a 
numerical solution of the GL equation
\begin{equation}
-{\hbar^2\nabla^2\over 2m^*}\psi+
{e^{*2}{\bf A}^2\over 2m^*}\psi-
{\gamma T_c^2\over 2n}\left(1-{t^2\over\sqrt{1-{2\over n}
\psi^2}}\right)\psi=0,
\label{a1}
\end{equation}
where $n$ is the density of pairable electrons, $\gamma$ is
the linear coefficient of the specific heat, $m^*=2m$ and 
$e^*=2e$ are the mass and charge of the Cooper pair, and
$t=T/T_c$ is the temperature on the scale of the critical
temperature. In the assumed geometry one can use the London
gauge in which the GL wave function is real.

Equation (\ref{a1}) has been proposed by Bardeen as an 
extension to low temperatures.\cite{B55} In fact, Bardeen 
arrived to this equation independently adding the quantum 
kinetic energy to the Gorter-Casimir two-fluid model. 
Close to the critical 
temperature the GL function has a small magnitude
$2\psi^2\ll n$ which allows one to expand the square-root
$1/\sqrt{1-{2\over n}\psi^2}\approx 1+{1\over n}\psi^2$,
and one finds the customary GL equation with parameters 
$\alpha=\gamma T_c^2(t^2-1)/2n$ and $\beta=\gamma 
T_c^2t^2/n^2$.

The vector potential $\bf A$ is given by the Maxwell equation
\begin{equation}
\nabla\times\nabla\times{\bf A}=\mu_0{e^{*2}\over m^*}
\psi^2 {\bf A}.
\label{a2}
\end{equation}
Equations (\ref{a1}) and (\ref{a2}) are solved together with
$\gamma$ and $T_c$ kept constant.

From the GL wave function $\psi$ we evaluate the Bernoulli
potential\cite{LKMB02}
\begin{eqnarray}
e\varphi-\lambda_{\rm TF}^2\nabla^2e\varphi&=&-
{\gamma T_c^2\over 2n^2}\left(1-{t^2\over\sqrt{1-{2\over n}
\psi^2}}\right)\psi^2
\nonumber\\
&&+{\psi^2\over 2n}{\partial\gamma T_c^2\over\partial n}+
{t^2T_c^2\over 2}{\partial\gamma\over\partial n}\sqrt{1-
{2\over n}\psi^2}.
\label{a3}
\end{eqnarray}
The density dependence of $\gamma$ and $T_c$ has been
estimated from the experimental specific heat, theoretical
free-electron density of states, and the McMillan 
formula\cite{KSK74},
see Appendix in Ref.~\onlinecite{LKMB02}. The Thomas-Fermi
screening length, 
$\lambda_{\rm TF}^2={\pi^2k_B^2\epsilon_0\over\gamma 
e^2}$,
is very small and its contribution can be
neglected, because the general integral which is proportional to $\exp(-x/
\lambda_{\rm TF})$ is zero since no surface charge forms on 
this scale.

In the final step the charge profile is evaluated from the
Poisson equation (\ref{a4}).

\section{Low magnetic fields}
In weak magnetic fields the GL wave function is only little
perturbed from its bulk value $\psi=\psi_\infty+\delta\psi$
with $\delta\psi\ll\psi_\infty$. In this limit the charge 
profile can be solved analytically. 

In the Maxwell equation (\ref{a2}) the vector potential is
a small quantity so that the perturbation of the wave does not
contribute in lowest order. The vector potential can thus 
be solved taking $\psi\approx \psi_\infty=\sqrt{n(1-t^4)/2}$. 
For both types of superconductors, this results in a simple 
exponential decay of the vector potential,
\begin{equation}
{A}_x\approx B\lambda{\rm e}^{-x/\lambda}.
\label{a6}
\end{equation}
on the scale of the London penetration depth $\lambda=\lambda_0/
\sqrt{1-t^4}$, with $\lambda_0^2=2m^*/(\mu_0e^{*2}n)$. Here 
$B$ is the value of the magnetic field at the surface at $x=0$.

\subsection{GL wave function}
The linear approximation of the GL equation (\ref{a1}) reads
\begin{equation}
{\hbar^2\nabla^2\over 2m^*}\delta\psi-
{\gamma T_c^2t^2\over n^2}
\left(1-{2\over n}\psi^2_\infty\right)^{-{3\over 2}}
\psi_\infty^2\delta\psi=
{e^{*2}{\bf A}^2\over 2m^*}\psi_\infty,
\label{a7}
\end{equation}
which can be expressed in terms of the GL coherence length 
$\xi^2=2\hbar^2 nt^4/(m^*\gamma T_c^2(1-t^4))$ as
\begin{equation}
\nabla^2\delta\psi-{2\over\xi^2}\delta\psi={e^{*2}\over\hbar^2}
\psi_\infty \lambda^2B^2{\rm e}^{-{2x/\lambda}}.
\label{a8}
\end{equation}

The solution of (\ref{a8}) is composed from the particular integral
with the decay on the scale of the London penetration depth 
and the general integral decaying on the scale of the GL 
coherence length,
\begin{equation}
\delta\psi=\psi_\lambda{\rm e}^{-{2x/\lambda}}+
\psi_\xi{\rm e}^{-{\sqrt{2}x/\xi}}.
\label{a9}
\end{equation}
The general integral of (\ref{a8}) can also include a growing 
term $\propto\exp(\sqrt{2}x/\xi)$. This term is excluded, 
because the perturbation asymptotically vanishes in the bulk, 
$\delta\psi\to 0$ for $x\to\infty$.

From the GL boundary condition, $\nabla\delta\psi=0$ at 
$x=0$, we find that the two amplitudes are linked by
\begin{equation}
\psi_\xi=-{\sqrt{2}\over\kappa}\psi_\lambda,
\label{a10}
\end{equation}
where $\kappa=\lambda/\xi$. We note that within Bardeen's 
extension of the GL theory, the ratio $\lambda/\xi$ depends 
on the temperature, $\kappa=\kappa_{\rm GL}/t^2$, where
$\kappa_{\rm GL}={m^*T_c\over ne^*\hbar}\sqrt{\gamma\over
\mu_0}$ is the ordinary GL parameter defined at $T_c$. 
The diverging effective GL parameter at low temperatures 
shows that within Bardeen's approximation any system 
behaves as a type-II superconductor at sufficiently low 
temperatures. This feature contrasts with the linearly 
decreasing effective GL parameter one finds close to the upper 
critical magnetic field.\cite{LKMB02}. Of 
course, the strong temperature dependence of $\kappa$ is
rather an artifact of the model. Since $\kappa$ is the most
important quantity of the GL theory, one can simply take 
$\lambda$ and $\kappa$ as input parameters.

The amplitude of the particular integral is obtained from the GL
equation (\ref{a8}) as
\begin{equation}
\psi_\lambda=\psi_\infty {{e^{*2}\over\hbar^2}\lambda^2B^2
\over{4\over\lambda^2}-{2\over\xi^2}}=
\psi_\infty 
{e^{*2}\lambda^4B^2\over 2\hbar^2}{1\over 2-\kappa^2}.
\label{a11}
\end{equation}
The GL wave function from (\ref{a9}-\ref{a11}) results in
\begin{equation}
\delta\psi={\psi_\infty\over 2-\kappa^2}
{e^{*2}\lambda^4B^2\over 2\hbar^2}\left(
{\rm e}^{-{2x/\lambda}}-{\sqrt{2}\over\kappa}
{\rm e}^{-{\sqrt{2}x/\xi}}\right).
\label{a12}
\end{equation}

\subsection{Bernoulli potential}
Finally, we take the linear approximation of the Bernoulli 
potential (\ref{a3}) in $\delta\psi$. Since the asymptotic value 
of the potential $\varphi_\infty=\varphi[|\psi_\infty|^2]$ is not 
essential, we focus on the perturbation caused by the magnetic
field
\begin{equation}
e\delta\varphi-\lambda_{\rm TF}^2\nabla^2e\delta\varphi=C
\delta\psi~.
\label{a13}
\end{equation}
In terms of $\lambda$ and $\kappa$ the coefficient $C$ reads
\begin{equation}
C={\hbar^2\kappa^2\over 2m^*\lambda^2\psi_\infty}
\left(1-t^4+4t^4{\partial\ln T_c\over\partial\ln n}\right).
\label{a14}
\end{equation}
A general integral which decays into the bulk is of the form
\begin{equation}
e\delta\varphi=e\delta\varphi_{\rm TF}{\rm e}^{-
{x/\lambda_{\rm TF}}}+C_\lambda\psi_\lambda
{\rm e}^{-{2x/\lambda}}+C_\xi\psi_\xi
{\rm e}^{-{\sqrt{2}x/\xi}}.
\label{a15}
\end{equation}
From (\ref{a13}) one finds the amplitudes
\begin{eqnarray}
C_\lambda&=&
{C\over 1-4{\lambda_{\rm TF}^2\over\lambda^2}},
\label{a16}\\
C_\xi&=&{C\over 1-2{\lambda_{\rm TF}^2\over\xi^2}}.
\label{a17}
\end{eqnarray}

The amplitude of the general integral is given by the charge 
neutrality, which requires $\nabla\delta\varphi=0$ at $x=0$.
From (\ref{a15}) follows
\begin{equation}
e\delta\varphi_{\rm TF}=-2{\lambda_{\rm TF}\over\lambda}
C_\lambda\psi_\lambda
-\sqrt{2}{\lambda_{\rm TF}\over\xi}C_\xi\psi_\xi.
\label{a18}
\end{equation}
The relative amplitudes proportional to $\lambda_{\rm TF}/\lambda$ and
$\lambda_{\rm TF}/\xi$ are not sufficiently small to be 
neglected. The charge density is given by the second 
derivatives so that one obtains relative amplitudes of the 
charge density at the surface proportional to $\lambda/\lambda_{\rm TF}$ 
or to $\xi/\lambda_{\rm TF}$. The surface charge is an 
integral, i.e. the contribution in question is proportional to the 
amplitude at the surface multiplied with the Thomas-Fermi length.
According to condition (\ref{a18}) all charges can have 
comparable values.

To prove that the screening on the scale of the Thomas-Fermi length $\lambda_{\rm TF}$ and the related surface charge can be neglected, one has 
to employ the GL boundary condition $\nabla\delta\psi=0$ at 
$x=0$. Using (\ref{a10}) one obtains
\begin{equation}
e\delta\varphi_{\rm TF}=-{\lambda_{\rm TF}^3\over\lambda^3}
{4(2-\kappa^2)\over1-2\kappa^2{\lambda_{\rm TF}^2/
\lambda^2}}C_\lambda\psi_\lambda
\label{a19}
\end{equation}
so that the general integral is proportional to the cube of the small
ratio $\lambda_{\rm TF}/\lambda$. Accordingly, the charge
due to the general integral is smaller by a factor $\lambda_{\rm TF}^2/\lambda^2$ 
than the charge created by the particular integral. In the 
following we neglect this contribution for simplicity of notation. 
Within the same level of accuracy we take $C_\lambda=C$ and 
$C_\xi=C$.

Now we are ready to evaluate the electrostatic potential. Using
(\ref{a10}) and (\ref{a11})  we obtain from (\ref{a15})
\begin{eqnarray}
e\delta\varphi&=&-{B^2\over 2\mu_0 n}
\left(1+{4t^4\over 1-t^4}{\partial\ln T_c\over\partial\ln n}\right)
\nonumber\\
&&\times{1\over 1-{2\over\kappa^2}}
\left({\rm e}^{-{2x/\lambda}}-{\sqrt{2}\over\kappa}
{\rm e}^{-{\sqrt{2}x/\xi}}\right).
\label{a20}
\end{eqnarray}
This linearized relation extends Rickayzen's formula to type-I
and weak type-II superconductors.

Let us first take a look at the extreme type-II superconductor, $\kappa\gg 1$, for
which Rickayzen's formula is recovered. For the assumed real 
GL wave function, the current is proportional to the vector 
potential ${\bf j}={e^{*2}\over m^*}\psi_\infty^2{\bf A}={e^2\over 
m}n_s{\bf A}$. With the velocity defined via the current, ${\bf j}
=e^*\psi_\infty^2{\bf v}=en_s{\bf v}$, one finds from (\ref{a6}) the relation 
between the magnetic pressure and the kinetic energy ${\rm e}^{-2x/\lambda}{B^2\over 2\mu_0 n}={n_s\over n}{1\over 2}mv^2$. Equation (\ref{a20}) yields then
the Bernoulli potential as
\begin{eqnarray}
e\delta\varphi&=&-{1\over 2}mv^2
\left({n_s\over n}+4{n_n\over n}{\partial\ln T_c\over\partial
\ln n}\right)
\nonumber\\
&&\times{1\over 1+{\sqrt{2}\over\kappa}}\left(
1+{{\rm e}^{\left(1-{\kappa/\sqrt{2}}\right){2x/\lambda}}-
1\over 1-{\kappa\over\sqrt{2}}}\right).
\label{a21}
\end{eqnarray}
We have used $n_s=n(1-t^4)$ and $n_n=nt^4$ to express the 
temperature dependence in terms of the condensate fraction.

One can see that Rickayzen's formula (\ref{waldram}) holds for extreme type-II
superconductors, $\kappa\to\infty$. In this limit the general integral
has a vanishingly small amplitude and the factor $1/(1+{\sqrt{2}
\over\kappa})$ goes to unity. For finite $\kappa$ the general  integral deforms the profile of the potential and the particular integral
has a reduced amplitude. 

Within Bardeen's extension of the GL theory, Rickayzen's
formula is always recovered at low temperatures as $\kappa$ 
diverges with $1/t^2$. Since Rickayzen's formula approaches the 
plain Bernoulli potential for $t\to 0$, the presented derivation also 
reproduces London's result.

\subsection{Behavior for finite $\kappa$}

As already shown, for the extreme type-II superconductor, 
$\kappa\to\infty$, the general integral vanishes as $1/\kappa$ and 
far from the surface it decays faster than the particular integral. 
The Bernoulli potential thus extends on the scale $\lambda/2$.

For the finite GL parameter in the region $\kappa>\sqrt{2}$ the
overall picture is rather similar. The general integral is appreciable
only close to the surface so that the scale on which the Bernoulli
potential extends is still $\lambda/2$. Besides, the amplitude of 
the particular integral is enhanced by the factor $1/(1-2/\kappa^2)$.

For the region $\kappa<\sqrt{2}$, the role of the general and
particular integrals are reversed. First of all, the amplitude of 
the general integral is larger than the amplitude of the particular integral. 
Since the general integral also decays more slowly in this case, 
the Bernoulli potential is dominated by the general integral and 
extends on the scale $\xi/\sqrt{2}$. The reversed role is also
supported by the fact that the signs of both parts are opposite as
compared to the case $\kappa>\sqrt{2}$. This sign reversal 
appears due to the enhancement factor $1/(1-2/\kappa^2)$.

Apparently, both components of $\delta\varphi$
diverge as $\kappa$ approaches $\sqrt{2}$. Their sum 
remains regular, however. For this particular case the scale of
both contributions is identical, namely $\lambda/2$. It is
advantageous to use expression (\ref{a20}) to obtain the 
asymptotic form of the Bernoulli potential 
\begin{equation}
\lim\limits_{\kappa\to\sqrt{2}} \,\,e\delta\varphi= -{m\over 4}v^2
\left({n_s\over n}+4{n_n\over n}{\partial\ln T_c\over\partial
\ln n}\right)\left(1+{2x\over\lambda}\right).
\label{a22}
\end{equation}

\begin{figure}[h]  % F2
\psfig{figure=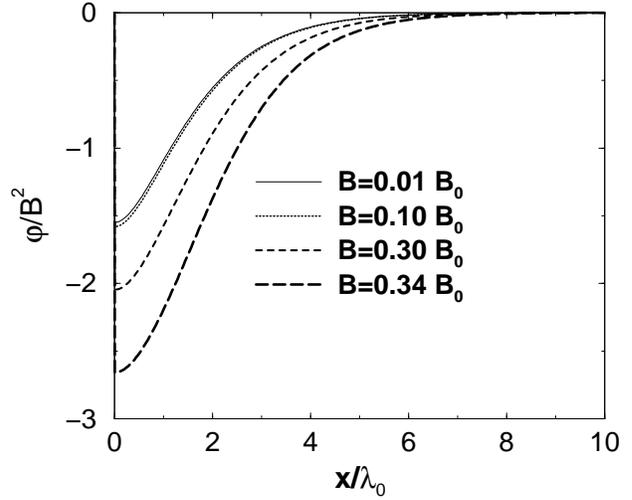,width=8cm}
\caption{Bernoulli potential as a function of the distance from the surface for 4 different magnetic 
fields. The parameters are the same as in Fig.~\ref{f1}.}
\label{f2}
\end{figure}

The Bernoulli potentials for different values of the magnetic field
are shown in Fig.~\ref{f2}. The limit of low magnetic fields agrees 
with the analytical formula (\protect\ref{a21}), of course. 
One can see that for 
the two lowest fields $B=0.01$ and $B=0.1$, the potential keeps 
its profile while its magnitude scales with $B^2$. For magnetic 
fields close to the critical value $B_{c2}\approx 0.34$, the 
linear theory does not apply. 

\subsection{Charge profile}
The charge density results from the Bernoulli potential 
(\ref{a20}) and the Poisson equation (\ref{a4}) as
\begin{equation}
\rho={2e\epsilon_0 B^2\over m\left(1\!-\!{2\over\kappa^2}\right)}
\left({n_s\over n}\!+\!4{n_n\over n}
{\partial\ln T_c\over\partial\ln n}\right)\!\!
\left({\rm e}^{\!-\!{2x/\lambda}}\!-\!{\kappa\over\sqrt{2}}
{\rm e}^{\!-\!{\sqrt{2}x/\xi}}\right)\!.
\label{a23}
\end{equation}
Note that the relative amplitude of the two contributions is 
reciprocal to the relative amplitude of the potentials. Integrating
from zero to infinity one can check that the total charge is zero,
as it is required by the charge neutrality.

The resulting charge density is small compared to the total
density of electrons $en$. The ratio of the amplitude of (\ref{a23}) to the total electron density can be written as
\begin{equation}
{\rho \over e n}\approx {\epsilon_0 B^2\over m n}={B^2\over \mu_0 n}{1\over mc^2}.
\label{a24}
\end{equation}
Even for upper estimates of the magnetic field by the critical 
value, $B\approx B_c$, the factor (\ref{a24}) is extremely small. For 
niobium  $B^2/\mu_0={1\over 2}\gamma T_c^2=3.2\times10^4
$~J/m$^3$, while $mc^2\approx 10^{-13}$~J. This suggests 
that the maximum charge density will be of the order of 
$10^{18}$ electrons per cubic meter. This is by ten orders of
magnitude smaller than the electron density of niobium $n=2.2
\times 10^{28}$~m$^{-3}$. This small value of the accumulated or 
depleted charge justifies to ignore the charge profile in the 
Maxwell equation (\ref{a2}) and the change of material 
parameters like $\gamma$, $n$ and $T_c$ in the GL equation 
(\ref{a1}).

Similarly to the potential, the charge profile reflects two scales,
$\lambda/2$ and $\xi/\sqrt{2}$. Before we analyze its properties
for various values of the GL parameter $\kappa$, we want to
discuss a few general features. First, the surface value of the
charge, obtained for $x=0$ in (\ref{a23}), is always negative, i.e. the charge carriers are 
depleted. We call the layer of the negative charge the surface
charge. Second, in the bulk sufficiently far from the surface the 
charge is positive, i.e., the charge carriers are accumulated. 
We call the region of positive charge the bulk charge. Third,
the width $w$ of the surface is given by $\rho(w)=0$. From
(\ref{a23}) follows 
\begin{equation}
w={\lambda\over 2}{\ln{\kappa\over\sqrt{2}}\over
{\kappa\over\sqrt{2}}-1}.
\label{a25}
\end{equation}

In the extreme type-II superconductor, $\kappa\to\infty$, the 
surface charge is formed by the contribution on the scale of the 
GL coherence length $\xi$. The width of the surface goes to
a value $w\to {\xi\over\sqrt{2}}\ln{\kappa\over\sqrt{2}}$. The
bulk charge extends on the scale of the London penetration
depth as it is known from the classical picture.

For the limiting case $\kappa\to\sqrt{2}$ the width of the 
surface charge is $\lambda/2$. The charge density has the
profile 
\begin{equation}
\rho=-{e\epsilon_0B^2\over m} \left({n_s\over n}\!+\!4{n_n\over n}
{\partial\ln T_c\over\partial\ln n}\right){\rm e}^{-{2x/\lambda}}
\left(1-{2x\over\lambda}\right).
\label{a26}
\end{equation}
Note the different sign inside the last parentheses compared 
to the Bernoulli potential (\ref{a22}). 
Since $\lambda/2=\xi/\sqrt{2}$ one cannot associate the 
surface and the bulk charge to the GL coherence length $\xi$ or 
to $\lambda$.

For $\kappa<\sqrt{2}$, the surface charge is formed by the
contribution on the scale of the London penetration depth. For $\kappa\ll
\sqrt{2}$, the width of the surface charge is $w\to {\lambda
\over 2}\ln{\sqrt{2}\over\kappa}$. The bulk charge extends on
the scale of the GL coherence length $\xi$ as one can see from (\ref{a23}).

\section{Strong magnetic field}
In strong magnetic fields the GL wave function is 
suppressed in the vicinity of the surface which 
has to be accounted for in the Maxwell equation. In this case
one has to face the fact that the system of equations
(\ref{a1}) and (\ref{a2}) is nonlinear. We present only a few numerical
solutions to point out some features typical for strong magnetic fields.

\begin{figure}[h]  % F3
\psfig{figure=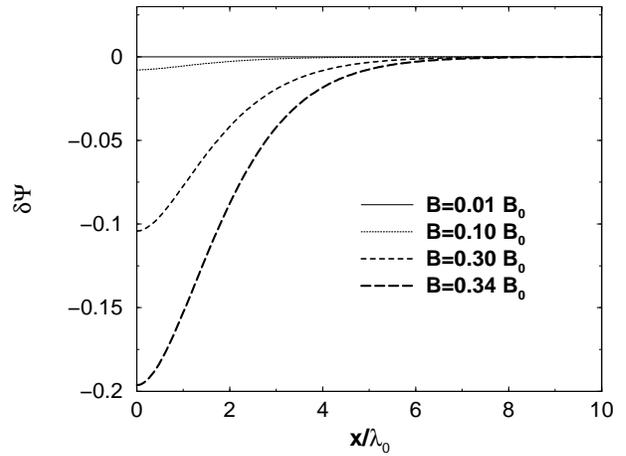,width=8cm,angle=-90}
\caption{The deviation of the wave function from its bulk value
$\psi_\infty=\sqrt{n(1-t^4)/2}\approx 0.6$ as a function of the distance from the surface. 
The parameters are the same as in Fig.~\ref{f1}.}
\label{f3}
\end{figure}

The nonlinear effects in strong magnetic fields follow from
a deviation of the wave function from its bulk value. In the
dimensionless representation used in Fig.~\ref{f3} the bulk
value is $\psi_\infty=\sqrt{n/2}\sqrt{1-t^4}\approx 0.6$. For the lowest
field $B=0.01$, the deviation is nearly zero on the scale of the 
graph. Also for $B=0.1$ the approximation $|\delta\psi|\ll
\psi_\infty$ is justified. For higher fields the deviation
becomes appreciable.

The suppression of the wave function at the surface shown in
Fig.~\ref{f3} affects the Bernoulli potential by two different 
mechanisms. First, the screening current is reduced at the 
surface so that the magnetic field penetrates deeper into the 
superconductor. This feature can be seen in Fig.~2, 
however, it is better visible in the profile of the charge 
density shown in Fig.~\ref{f1}, as a widening of the surface layer.

\begin{figure}[h]  % F4
\psfig{figure=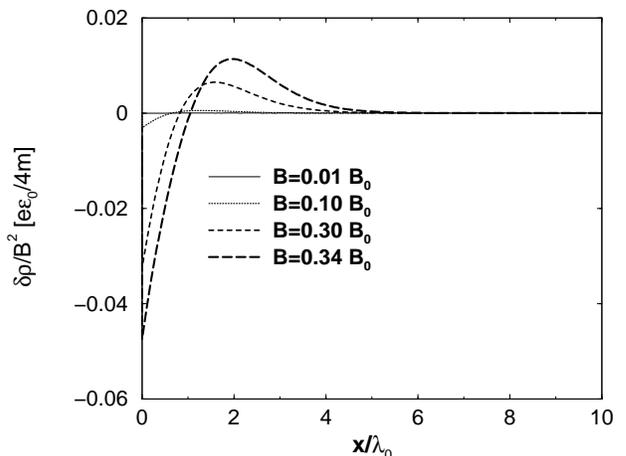,width=8cm,angle=-90}
\caption{The charge density which is proportional to the density
dependence of the linear coefficient of the specific heat versus distance from the surface.
We use the value for niobium, $\partial\ln\gamma/\partial\ln 
n=0.42$, which is close to the parabolic band approximation
$\gamma_{\rm par}\propto n^{1/3}$, i.e., 
$\partial\ln\gamma_{\rm par}/\partial\ln n=1/3$.}
\label{f4}
\end{figure}

Second, the thermodynamic contributions to the Bernoulli 
potential are nonlinear in the wave function, see (\ref{a3}).
In contrast, the linearized approximation of the Bernoulli
potential (\ref{a20}) or (\ref{a21}) does not depend on the
density derivative of the linear coefficient of the specific
heat, $\partial\gamma/\partial n$, included in the derivative
of the condensation energy ${1\over 4}\gamma T_c^2$. Beyond 
the linearized approximation, this derivative contributes as
demonstrated in Fig.~\ref{f4} for parameters of niobium. 

Comparing the contribution of $\partial\gamma/\partial n$ 
with the total charge density shown in Fig.~\ref{f1}, one 
can see that this particular nonlinear mechanism is rather 
small giving less than 3\% of the total charge 
density. This small value justifies the 
approximations\cite{KK92,KF95,B96} using only the term 
proportional to $\partial T_c/\partial n$.

\section{Conclusions}
In conclusion, we have shown that the Bernoulli potential in 
superconductors can be discussed by the charge build up in the 
region between the surface and $max[\xi/\sqrt{2},\lambda/2]$. The thinner depleted charge 
region carries the surface charge. In contrast to former 
theories which assumed that the surface charge is localized on 
the scale of Thomas-Fermi screening length $\lambda_{\rm TF}$, 
we find that the surface charge extends over a range 
$L=min[\xi/\sqrt{2},\lambda/2]$. In fact, there is a nonzero 
contribution on the scale of $\lambda_{\rm TF}$, but 
of a negligibly small amplitude proportional to $\lambda_{\rm TF}^2/L^2$.

For extreme type-II superconductors we have confirmed 
the picture known already from London with the thermodynamic 
corrections by Rickayzen. The bulk charge extends on the scale 
of the London penetration depth $\lambda/2$.
The surface charge is 
localized on the scale of the GL coherence length $\xi/\sqrt{2}$ 
which is negligible in this limit anyway. This is in agreement 
with the classical treatment 
which neglects all gradient contributions. 

In type-I or weak type-II superconductors, i.e for the GL parameter 
$\kappa<\sqrt{2}$, one finds the opposite situation. The surface 
charge is localized on the scale of the London penetration 
depth $\lambda$ while the bulk extends on the scale of the GL coherence length 
$\xi/\sqrt{2}$. In 
this case the gradient contributions are dominant and local
London-type theories naturally fail.

In strong magnetic fields the Bernoulli potential becomes a 
non-quadratic function of the magnetic field. This feature
parallels the nonlinear susceptibility as it is dominated
by the enhanced penetration of strong magnetic fields into
superconductors.
There is an additional thermodynamic correction absent in low 
magnetic fields, however, for realistic material parameters 
(like for niobium) it accounts for less than 3\% of the charge density. 
Thus the 
approximations used in \cite{KK92,KF95,B96} are justified.

Finally we remind that the potential $\phi$ discussed here 
does not include the surface dipole which affects the potential 
seen outside the sample. We will discuss this contribution in a forthcoming 
paper.

\acknowledgements
This work was supported by M\v{S}MT program Kontakt 
ME601 and GA\v{C}R 202/03/0410, GAAV A1010312 grants. 
The European (ESF) program VORTEX is also acknowledged.

\end{document}